\documentclass[prb,twocolumn,floatfix,showpacs]{revtex4-1}
\usepackage{graphicx,epsfig,color}

\begin{document}

\title{Range separated hybrid exchange-correlation functional analyses of W and/or N(S)
(co)doped anatase TiO$_2$}

\author{ Veysel \c{C}elik and  
Ersen Mete\footnote{Corresponding author: \indent e-mail:
emete@balikesir.edu.tr} }

\affiliation{Department of Physics, Bal{\i}kesir University,
Bal{\i}kesir 10145, Turkey}

\date{\today}

\begin{abstract}
Electronic properties and atomic structures of W, N, S, W/N, and W/S dopings 
of anatase TiO$_2$ have been systematically investigated using the density 
functional theory (DFT). The exchange and correlation effects have been treated 
with Heyd, Scuseria and Ernzerhof (HSE) hybrid functional. Mixing traditional 
semi-local and non-local screened Hartree-Fock (HF) exchange energies, the HSE 
method corrects the band gap and also improves the description of anion/cation 
derived gap states. Enhanced charge carrier dynamics, observed for W/N codoped 
titania, can partly be explained by the passivative modifications of N 2$p$ and 
W 5$d$ states on its electronic structure. Following this trend we have found an 
apparent band gap narrowing of 1.03 eV for W/S codoping. This is due to the large 
dispersion of S 3$p$ states at the valance band (VB) top extending its edge to 
higher energies and Ti--S--W hybridized states appearing at the bottom of the 
conduction band (CB). W/S-TiO$_2$ might show strong visible light response 
comparable to W/N codoped anatase catalysts.
\end{abstract}

\pacs{71.20.Nr, 71.55.-i, 61.72.Bb}

\maketitle

\section{Introduction}

In the growing area of renewable energy, TiO$_2$ (titania) is the most 
appropriate metal oxide for photocatalytic processes due to its powerful 
oxidation and charge transport properties along with its abundance, 
nontoxicity, and stability to corrosion. It is promising in applications 
such as photogeneration of hydrogen from water, dye sensitized solar cells 
(DSSC), degradation of pollutants under visible light irradiation and 
production of hydrocarbon fuels.\cite{Fujishima,Gratzel,Khan,Varghese}
Particularly, anatase phase has received an increased attention for it exhibits 
higher catalytic activity relative to rutile and brookite.\cite{Xu} Intrinsic 
wide band gap of pure TiO$_2$ ($\sim$3.2 eV for anatase~\cite{Tang} and 
$\sim$3.0 eV for rutile~\cite{Pascaul}) confines its photon absorption to 
ultraviolet (UV) region severely limiting solar energy utilization to $\sim$5\%. 
Great efforts have been made to modify the electronic properties of titania in 
order to extend its optical absorption edge into visible region and enhance its 
photoresponse. For this purpose, numerous studies have proposed doping of 
TiO$_2 $ with substitutional cations and/or anions as an effective 
approach.~\cite{Mu,Choi,Sakthivel,Sun,Wang,HYu,Zhu,Long,Yamamoto,Yin}

Nitrogen doped titania is considered to be one of the most effective 
photocatalysts. Although it has been extensively studied by both experimental 
and theoretical methods,\cite{Asahi,Irie,Spadavecchia,Katoh,Yamanaka,Gole,Nakano,Livraghi,Tao,Harb} 
proper description of midgap states is still desirable. For instance, 
Asahi~\textit{et al.}\cite{Asahi} concluded that N substitution for O causes 
a band gap narrowing due to mixing of 2$p$ states of N and O. On the contrary, 
Irie \textit{et al.}\cite{Irie} have suggested that the visible-light response 
in N-doped titania may be due to N 2$p$ states isolated above the valence band 
maximum (VBM). The nature of N-induced modifications to the electronic band 
structure depends on the doping content.\cite{Gole, Spadavecchia} Then, another 
question relates to increasing charge trapping rate with increasing N-doping 
which is associated with a localized state below the CB.\cite{Katoh} Most of
the theoretical studies failed to put forward the existence of such a trap level.

Sulphur is another effective dopant in modifying the electronic structure of 
titania. S can be substituted either at an O site as an anion or at a Ti site 
as a cation, depending on the incorporation techniques.
\cite{Umebayashi,Umebayashi2,Ohno,JCYu,Ho,Li,Tian,KYang}
Both of them exhibit high photocatalytic activity under visible light.\cite{Li} 
Moreover, visible light absorbance increases with the atomic percentage 
of S-dopant.\cite{JCYu,Ho} A reliable explanation based on the effect of S 3$p$ 
states on the electronic structure of TiO$_2$ is desirable. 

Monodoped TiO$_2$ is usually inefficient for light harvesting. For instance,
increasing concentration levels of transition metal doping, such as W 
incorporation, has detrimental effects on photogenerated charge carrier 
mobility.\cite{Fuerte,Couselo,YYang} Well localized isolated energy levels 
might act as electron-hole recombination centers reducing photocatalytic 
efficiency. To passivate such trap states, $n$-$p$ type codoping with anion and 
cation pairs has been proposed.\cite{Sun,Wang,HYu} Suitably chosen codopants 
not only narrow the band gap, but also serve to counteract the presence of recombination 
centers.\cite{Gai,Long_WN1,Long_WN2} For this purpose, many experiments have 
recently focused on the additional substitution of W into N-doped titania and reported 
significant enhancement of catalytic activity under visible-light irradiation relative 
to monodoped reference systems. \cite{Gao,Shen,Kubacka,Kubacka1,Kubacka2}

We used screened exchange hybrid DFT method for a proper 
description of the electronic structure of N, S, W, W/N and W/S doped 
anatase TiO$_2$. We attempt to analyze densities of states (DOS) in 
relation to visible light absorbance and charge carrier trapping, by 
comparing with experimental observations where 
explanations are still needed. We also discuss atomic structures, 
thermodynamical energetics and charge states of these ions 
substituted into the anatase lattice.

\section{Computational Method}

The spin-polarized hybrid density functional theory calculations were carried 
out using the Vienna \textit{ab-initio} simulation package (VASP).\cite{vasp} 
Ionic cores and valence electrons were treated by projector-augmented waves 
(PAW) method.\cite{paw1,paw2} Plane wave basis set was used to expand the 
wavefunctions up to a kinetic energy cutoff value of 400 eV. We used fine 
FFT grids with high precision settings throughout the calculations. 

Standard DFT usually employs local or semilocal approximations to the 
exchange-correlation (XC) energy.  This leads to erroneous descriptions for 
some real systems like transition metal oxides. One way of overcoming this 
deficiency is to use hybrid functionals, where a portion of the non-local 
Hartree-Fock (HF) type exchange is admixed with a semilocal XC functional. We 
used the Heyd-Scuseria-Ernzerhof hybrid functional (HSE06)~\cite{Heyd1,Heyd2,Paier} 
which adopts a screened Coulomb potential. Hence, the exchange hole becomes 
delocalized around a reference point but not beyond, reducing self-interaction error. 
Moreover, resulting rapid spatial decay of HF exchange improves the convergence 
behavior of self-consistent procedures. The HSE exchange is derived from the 
PBE0\cite{PBE,Adamo} exchange by range separation and then by elimination of 
counteracting long range contributions, as,
\[
E_{x}^{\rm HSE}=aE_{x}^{\rm HF,SR}(\omega)+(1-a)E_{x}^{\rm PBE,SR}(\omega)+E_{x}^{\rm PBE,LR}(\omega),
\]
where $a$ is the mixing coefficient and $\omega$ is the range separation 
parameter. A consistent screening parameter of $\omega$=0.2 \AA$^{-1}$ is used 
for the semilocal PBE exchange as well as for the screened non-local exchange 
as suggested for the HSE06 functional.\cite{Krukau} The choice of exact exchange 
contribution coefficient is important. Becke \cite{Becke} derived a value of 20\% 
by fitting to atomization energy data of a large number of molecular species. 
Later, it was suggested to be 25\% by Perdew \textit{et al.}\cite{Perdew} 
We adjusted this mixing to be 22\% with which band gaps and lattice parameters 
for both anatase and rutile phases of titania can be obtained in good agreement 
with the experimental data as given in Table~\ref{table1}.

\begin{figure}[ht]
\epsfig{file=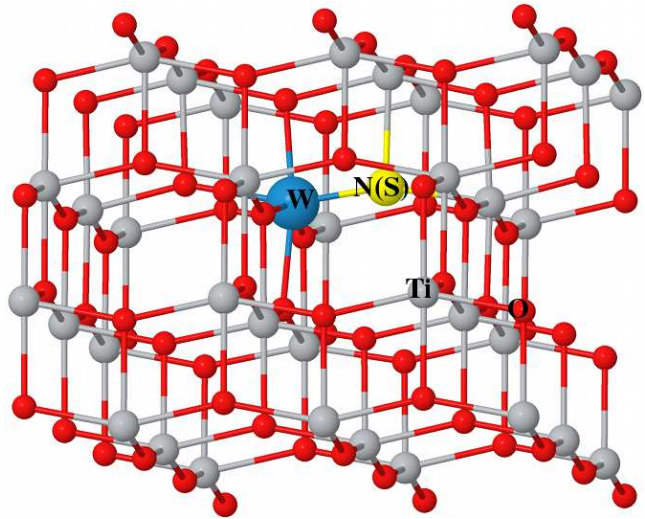,width=8.5cm}
\caption{Model bulk cell structure containing 108 atoms for W/N(S) (co)doped 
anatase TiO$_2$. Substitutional dopants are shown for N(S) at an O site, and 
for W at a Ti site.\label{fig1}}
\end{figure}

We used a 108-atom supercell constructed by 3$\times$3$\times$1 replication
of the anatase unit cell, which ensures sufficient spatial separation between 
the periodic images of the impurities as shown in Fig.~\ref{fig1}. Various dopings 
of TiO$_2$ have been modeled by substitution of N (or S)  at O and/or W at 
Ti sites.  We also tried S@Ti configuration and found that this is energetically 
1.246 eV/cell less favorable than S@O.  For geometry optimizations and 
electronic structure calculations, the Brillouin zone was sampled with 
2$\times$2$\times$2  mesh of special $k$-points. Previously, the same grid 
was shown to be sufficient to study oxygen vacancy formation in anatase 
TiO$_2$ using a smaller (76-atom)  cell.\cite{Janotti}  Atomic positions and 
cell parameters were fully optimized until spatial components of residual forces on 
each ion were below 0.015 eV/{\AA} performing collinear spin polarized calculations.

We calculated (co)dopant formation energies using,
\[
E_{f}=E_{\rm doped}-E_{\rm pure}-n\mu_{\rm W}-m\mu_{\rm N(S)}+n\mu_{\rm Ti}+m\mu_{\rm O},
\]
where E$_{\rm doped}$ and E$_{\rm pure}$ are the total energies of doped 
and pure supercells, while $\mu_{\rm W}$, $\mu_{\rm N}$, $\mu_{\rm S}$, 
$\mu_{\rm Ti}$ and $\mu_{\rm O}$ are the chemical potentials of the W, N, 
S, Ti and O species, respectively. The integer $n$ gives the number of W 
cations and $m$ denotes the number of N(S) anions. In thermodynamical 
equilibrium with the anatase phase, $\mu_{\rm Ti}$ and $\mu_{\rm O}$ must 
satisfy the relation $\mu_{\rm TiO_{2}}=\mu_{\rm Ti}+2\mu_{\rm O}$. The amount 
of Ti and O in a growth environment influences their chemical potentials. 
High(low) values of $\mu_{\rm O}$ correspond to O-rich(-poor) conditions and 
can also be interpreted as Ti-poor(-rich) conditions from the equilibrium 
relation. Under O-rich conditions, $\mu_{\rm O}$ is the half of the energy 
of an O$_2$ molecule $(E_{\rm O_2})$, and $\mu_{Ti}$ is obtained through 
the condition $\mu_{\rm Ti}=\mu_{\rm TiO_{2}}-E_{\rm O_2}$. Under Ti-rich 
conditions, $\mu_{\rm Ti}$ is the energy of a Ti atom in its bulk unit cell 
($\mu_{\rm Ti}^{\rm bulk}$) and $\mu_{\rm O}$ is calculated from the 
equilibrium restriction by $\mu_{\rm O}=\frac{1}{2}(\mu_{\rm TiO_{2}}-\mu_{\rm Ti})$. 
The chemical potentials of the dopants are extracted from their naturally 
occurring phases. $\mu_{\rm W}$ is calculated as $\mu_W=E_{\rm WO_3}-\frac{3}{2}E_{\rm O_2}$. 
Similarly, $\mu_{\rm S}$ is obtained from the relation $\mu_S=E_{\rm SO_2}-E_{\rm O_2}$. 
For N, we used $\mu_{\rm N}=\frac{1}{2}E_{\rm N_2}$. 
Calculated dopant formation energies are given in Table~\ref{table2}.

For the qualitative description of interatomic charge distributions, we used
Bader analysis based on atom in molecule (AIM) theory. Local charge 
depletion/accumulation can be computed by integrating Bader volumes
around atomic sites. These volumes are partitions of the real space cell 
delimited by local zero-flux surfaces of charge density gradient vector field.
We calculated charge states of atomic species (see Table~\ref{table3}) 
using a grid based decomposition algorithm developed by Henkelman's 
group.\cite{Henkelman}

\section{Results \& Discussion}

For the pure TiO$_2$, the calculated lattice parameters show remarkable 
agreement with the experimental data as presented in Table~\ref{table1}. The 
calculated band gap values using the bulk unit cells for both anatase and 
rutile polymorphs are largely corrected by the HSE functional. When we use 
108-atom supercell for the anatase case, it slightly changes to 3.23 eV as 
indicated in Fig.~\ref{fig4}a. The bottom of the CB is formed by highly 
dispersing Ti 3$d$. (see PDOS in Fig.~\ref{fig4}g) They are not isolated but 
a part of the CB in consistency with the energy bands of pure anatase 
calculated with HSE06 by Yamamoto~\textit{et al.}.\cite{Yamamoto}
Calculated Bader charges yield oxidation states of $+$2.84 and $-$1.43
for Ti and O, respectively. These are closer to formal values relative
to those obtained by PBE.\cite{Mete}

\begin{table}[h!]
\caption{Computational results and experimental data (in parantheses) for the bulk 
material properties of TiO$_2$. Calculations were performed using the HSE06 functional 
with a 22\% exact exchange contribution.\label{table1}}
\begin{ruledtabular}
\begin{tabular}{ccccc}
Phase & \multicolumn{2}{c}{Lattice parameters} &
Transition & Band gap \\[1mm]\hline
& a ({\AA})& c ({\AA}) & & (eV) \\[1mm] \hline
Rutile   & 4.57 (4.59)\cite{Burdett} ~ & 2.94 (2.95)\cite{Burdett} &$\Gamma\rightarrow\Gamma$& 2.97 (3.0)\cite{Pascaul} \\
Anatase & 3.78 (3.78)\cite{Burdett}~ & 9.45 (9.50)\cite{Burdett} &$Z\rightarrow\Gamma$& 3.20 (3.2)\cite{Tang} \\
\end{tabular}
\end{ruledtabular}
\end{table}

\begin{figure}[t]
\epsfig{file=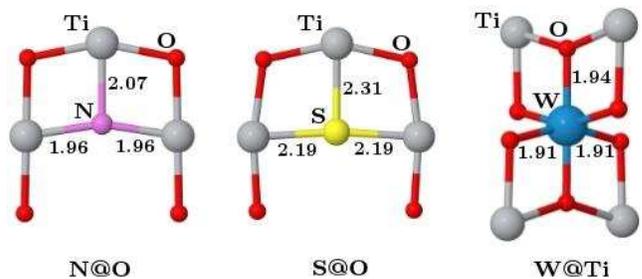,width=8.5cm}
\caption{Relaxed structures of monodopants, N, S and W inside anatase TiO$_2$ lattice.
Bond lengths are in angstroms.
\label{fig2}}
\end{figure}

\noindent\textbf{\emph{N@O doping :}}
Substitution of single N atoms at O sites was confirmed by several reports.
\cite{Asahi,Irie,Gole,Nakano} Relaxed geometry of N dopant in the crystal has 
been shown in Fig~\ref{fig2}. HSE functional gave a nearest neighbor Ti-N bond 
length of 1.96 {\AA} while Ti-O bond was found to be 1.93 {\AA} in its pure 
anatase phase. In addition, Ti-N bond along [001] is (2.07 {\AA}) slightly 
larger than that of the Ti-O being 1.97 {\AA}. Hence, N@O doping has a 
little effect on the lattice structure. Consistenly, the calculated charge 
states for N and O are $-$1.38$e$ and $-$1.43$e$, respectively. (see 
Table~\ref{table3}) Hence, replacement of an O with an N is not expected to 
cause significant distortions in the lattice structure. In fact, N@O has the 
lowest formation energy, 0.53 eV, under O-poor conditions among all considered 
in this study. (Table~\ref{table2}) When O is abundant in the environment it 
reasonably increases to 5.10 eV.

For the explanation of visible light response of N substituted TiO$_2$, the
experiments (other than the study Irie~\textit{et al.}\cite{Irie}) mainly agree on 
the band gap narrowing due to hybridization of N 2$p$ and O 2$p$ 
states.\cite{Asahi,Nakano,Livraghi,Katoh} This partial answer needs to be 
complemented with the possible presence of an impurity level below the CB 
as indicated by recombination characteristics of photogenerated electron-hole 
pairs\cite{Katoh} and variation of photoresponse under irradiation with different 
wavelengths.\cite{Irie,Livraghi,Spadavecchia}  Using X-ray photoelectron 
spectroscopy (XPS), Irie~\textit{et al.}\cite{Irie} suggested an isolated N 2$p$ 
narrow band formation above the VB, which is attributed to be responsible for the 
visible light sensitivity. Recently, deep-level optical spectroscopy (DLOS) 
measurements of Nakano~\textit{et al.} have located two deep levels at 
$\sim$1.18 and $\sim$2.48 eV below the CB.\cite{Nakano}  The 2.48 eV band 
has been identified to behave as a part of the VB  by mixing with the O 2$p$ 
bands,  which contributes to band gap narrowing. For the former, the possibility 
of oxygen vacancies depending on the processes involved in preparation of the 
samples were addressed as a potential origin of the impurity level closer to the CB.
\cite{Nakano,Livraghi,Katoh} However, this issue is still unclear. 
For instance, empty N orbitals might be involved.\cite{Spadavecchia}

From the theoretical side, the position and the number of defect states in the 
band gap show differences depending on the type of DFT methods. For instance, 
local density approximation (LDA) predicts single N 2$p$ state at the edge of 
VBM.\cite{Gai} Generalized gradient approximation (GGA) and GGA+U studies 
predicted it to be isolated above the VBM.\cite{Tao,Harb,Long_WN1,Long_WN2} In 
the DFT+U method, the supplementary on-site repulsion, U, acts only on Ti 3$d$ 
electrons and is specific to the system under consideration. In this study, we 
identified two N driven  gap states (in Fig.~\ref{fig4}b,\ref{fig4}h) based on 
the screened Coulomb potential approach. One of them is a singly occupied 
bonding state showing dominant N 2$p$ character and is at the edge of VBM 
causing a considerable band gap narrowing by mixing with O 2$p$ states. Moreover, 
PDOS reveals an increasing contribution of N toward the upper part of the VB. 
Consistently, the presence of single-atom N impurities as diamagnetic (singly 
occupied) centers was previously observed in the electron paramagnetic resonance 
(EPR) studies.\cite{Livraghi} The hybrid character of this state has been 
confirmed by many experiments.\cite{Asahi,Nakano,Livraghi,Spadavecchia,Katoh} 
This reflects a strong covalency of N centers with the TiO$_2$ matrix as shown 
in the ground state total electronic density in Fig.~\ref{fig5}a. From diffuse 
reflectance spectroscopy (DRS) measurements, Spadavecchia~\textit{et al.}
\cite{Spadavecchia} reported an ``apparent'' band gap of 2.88 eV at 0.4 N/Ti 
molar ratio, which matches with our calculations as depicted in 
Fig.~\ref{fig4}h. Similarly, Livraghi~\textit{et al.} found it to be 2.84 eV 
from their EPR spectra.\cite{Livraghi} 

\begin{table}[b!]
\caption{Calculated formation energies (eV) for W, S, N, W/N, W/S doping of anatase supercell
with 18 TiO$_2$ units.\label{table2}}
\begin{ruledtabular}
\begin{tabular}{cccccccc}
       &\multicolumn{1}{c}{Ti-rich}&\multicolumn{1}{c}{O-rich}\\[1mm]\hline
N-doped&0.53 & 5.10\\
W-doped&12.45&3.30\\
S-doped&3.58&8.16\\
W/N-doped&10.44&5.87 \\
W/S-doped&15.81&11.23\\
\end{tabular}
\end{ruledtabular}
\end{table}

Reports focusing on the charge carrier dynamics tend to offer the possibility 
of oxygen vacancy levels 0.75 -- 1.18 eV below the CB.\cite{Nakano,Katoh,Yamanaka} 
Since reduction of the charge separation efficiency under visible light 
excitation is significantly large, vacancies are believed to promote
recombination of holes and electrons. Our HSE calculations predicted an 
antibonding energy level for one spin component characterized dominantly 
by N 2$p$ isolated 0.63 eV below the CB (see Fig.~\ref{fig4}h). In the visible 
region, it might well act as a trap level centered at the N impurity site. The 
decrease in the quantum yields with increasing N concentration can also be 
associated to this empty N 2$p$ energy level which does not conflict with 
possible existence of adjacent oxygen vacancies. Consistently, 
Spadavecchia~\textit{et al.}\cite{Spadavecchia} opened up the possibility of 
charge transfer into empty N orbitals for the explanation of trapping mechanism.

\noindent\textbf{\emph{S@O doping :}}
Sulphur was shown to be incorporated into an O site of anatase lattice by 
oxidative annealing of TS$_2$.\cite{Umebayashi,Umebayashi2,Ho}  Cationic 
doping was also achieved by Ohno \textit{et al.}.\cite{Ohno} Recently, 
Li \textit{et al.} reported an alternative synthesis procedure for S 
substitution for O.\cite{Li} We relaxed both S@O and S@Ti configurations on
108-atom supercell and found that the former is energetically more favorable
by 1.246 eV/cell. Optimized geometry of a S dopant replacing the lattice O
is depicted in Fig.~\ref{fig2}. In comparison to an oxygen, the valence 
electrons of the S dopant have occupy more space. (see Fig.~\ref{fig5}b) 
Therefore, the Ti-S bond lengths become 2.19 and 2.31 {\AA.} while the 
corresponding Ti-O bonds are 1.93 and 1.97 {\AA} in undoped anatase. 
These local structural changes cause an increase in the formation energy 
(in Table~\ref{table2}) of an S dopant relative to that of N. The calculated 
value is 3.58 eV under O-poor conditions in reasonable agreement with a 
previous theoretical study.\cite{KYang} However, the charge state of S(O) 
is calculated with HSE as $-$1.17$e$($-$1.43$e$) (in Table~\ref{table3}) 
whereas it was previously predicted by pure DFT to be $-$0.28$e$ and  
$-$0.67$e$ for S and O, respectively.\cite{KYang} Nominally S is expected 
to remain as S$^{2-}$ similar to O$^{2-}$ in the crystal. Briefly,  the HSE 
functional improves the spatial distribution of charge densities  and 
estimates electronegativities closer to the nominal values.

\begin{table}[b!]
\caption{Average charge states ($e$) of dopants and their adjecent 
Ti and O atoms from Bader analysis.\label{table3}}
\begin{ruledtabular}
\begin{tabular}{cccccc}
Dopant& N      & S      & W     & Ti(nn)& O(nn)  \\[1mm]\hline
none  &        &        &       & +2.84 &$-$1.43 \\
N     &$-$1.38 &        &       & +2.31 &        \\
W     &        &        & +4.60 &       &$-$1.56 \\
S     &        &$-$1.17 &       & +2.78 &        \\
W/N   &$-$2.08 &        & +4.76 & +2.82 &$-$1.53 \\
W/S   &        &$-$1.29 & +4.20 & +2.73 &$-$1.54 \\
\end{tabular}
\end{ruledtabular}
\end{table}

Furthermore, screened hybrid functional better describes the effects of the S 
anion on the electronic structure of TiO$_2$ where pure DFT fails.  For 
instance, the PBE functional estimates a band gap narrowing of 0.45 eV at 
an S concentration of 0.0139 (one S atom in a 72-atom cell) after applying 
a scissors correction of 1.4 eV.\cite{Tian} Our HSE calculations estimate the 
apparent band gap as 2.24 eV that corresponds to an effective narrowing of 
0.99 eV for S@O doping in agreement with UV-vis diffuse reflectance spectra 
(DRS).\cite{Ho,Li} As seen in Fig.~\ref{fig4}i, a group of hybrid states localized 
at the edge of the VB resulting from O 2$p$, Ti 3$d$ and dominantly S 3$p$ 
coupling. This hybridization agrees well with the formation of S-Ti-O bonds 
as reported by Li~\textit{et al.} These hybrid states facilitate a 
state-to-band transition at 2.24 eV ($\sim$554 nm)  might explain the
observed absorbance edge at around 560 nm.\cite{Ho,Li} The increase in 
the visible DRS intensities with S concentration can be attributed to
increased excitation probabilities by larger DOS contribution and 
extended dispersion of S-Ti-O states.  In addition, the mixing of S 3$p$ with 
O 2$p$ band states increases VB width by 0.16 eV relative to the VBM 
of pure TiO$_2$.  Similarly, overlap of antibonding S 3$p$ orbitals with 
Ti 3$d$ states decreases the CBM by 0.13 eV. These changes in the band 
edges allow band-to-band transitions starting from 2.94 eV ($\sim$422 nm) .
This might be useful in explaining  the redshift of absorbance shoulder 
near UV in DRS data.\cite{Umebayashi,Ho,Li}

\noindent\textbf{\emph{W@Ti doping :}}
Experiments use sol-gel and microemulsion techniques to incorporate W into Ti 
sites.\cite{Fuerte,YYang} Hence, oxygen is always present in the environment. 
In parallel, we calculated a moderate formation energy of 3.30 eV for W@Ti 
doping under O-rich conditions. (see Table~\ref{table2}) On the other hand, 
when Ti is abundant, it gets as large as 12.45 eV. Substitution of W at Ti in 
the anatase lattice results in sixfold coordination as shown in Fig.~\ref{fig2}. 
Four conjugate W-O bonds are 1.91 {\AA} while the bonds along [001] direction 
become 1.94 {\AA}. These are slightly shortened relative to those of the 
undoped TiO$_2$. In fact, the ionicity of W-O bonding is lower then a Ti-O 
bond as seen in Fig.~\ref{fig5}c. In parallel, the Bader charges around 
W$^{6+}$ and Ti$^{4+}$ nominal species are found to be $-$1.40$e$ (W@Ti) 
and $-$1.16$e$ (Ti@pure), respectively. The disturbance on the lattice caused 
by the substitutional W is small and remains local. Although these values are 
lower than the formal charge states, partial inclusion of exact HF exchange 
improves them with respect to previous DFT results.\cite{Long_WN1} 

In addition, W-doping causes formation of occupied states 0.25 eV below the CB 
with W 5$d$ and Ti 3$d$ contributions (Fig.~\ref{fig4}d), which manifest similar 
DOS characteristics with an oxygen vacancy state\cite{Janotti,Mete} and also
with those of Nb and Ta monodopings described by Yamamoto~\textit{et al.}\cite{Yamamoto} 
Removal of an oxygen leaves excess electrons which accumulate around Ti ions.
In a similar manner, W@Ti introduces excess electrons, a portion of them 
accumulating around the W center. (Fig.~\ref{fig5}c) In fact, the oxidation
states of W and Ti species in Table~\ref{table3} reveal that the W valence shell 
5d$^4$6s$^2$ loses 4.60 of its charge while Ti 3d$^2$4s$^2$ gives 2.84 of it to 
form bonds with adjacent oxygens. Therefore, the charge accumulation around the 
W center amounts to 1.40$e$ being larger by 0.24$e$ than that of Ti. Moreover, a 
very weak coupling between W and Ti $d$ states can be seen in the PDOS analysis 
of these shallow states (Fig.~\ref{fig4}j). The W 5$d$ appears as a sharp dos peak 
for the majority spin component while Ti 3$d$ contribution is broader involving 
both spins. The existence of impurity levels close to the CB was also discussed
by experimenttal studies.\cite{Fuerte,YYang} Yang~\textit{et al.} reported a 
redshifted onset of optical absorption corresponding to an effective band 
gap of 2.73 eV for 3\% W doping. In parallel, HSE predicts the defect states 
lying 2.81 eV above the VBM. Under visible light irradiation 19 at.\% 
W-doped TiO$_2$ absorbs up to $\sim$445 nm (Fig.6 in Ref[\cite{Fuerte}])
that is in good agreement with our estimation of 442 nm. Further, increasing 
W content reduces photocatalytic activity.\cite{YYang} This can be ascribed 
to increased W 5$d$ contribution on the defect states which can act as trap 
levels.

\noindent\textbf{\emph{W/N codoping :}}
In an attempt to address the charge recombination bottleneck observed in 
N-doped TiO$_2$, Gao~\textit{et al.} proposed an additional coupling to 
WO$_3$.\cite{Gao} Numerous studies have shown drastic enhancement of 
the photocatalytic performance of titania based on passivation of the trapping 
mechanism by W-N codoping.\cite{Shen,Gao,Kubacka,Kubacka1,Kubacka2}

\begin{figure}[t!]
\epsfig{file=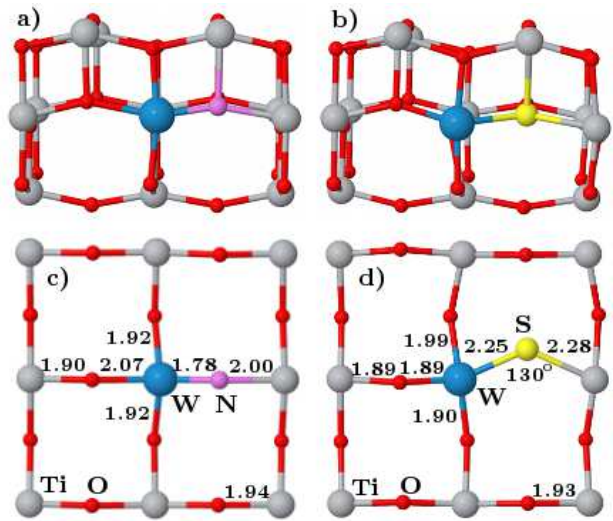,width=8.2cm}
\caption{Optimized atomic structures of W/N(S) codopants in anatase TiO$_2$.
Side views (a),(b) through [010] and top views (c),(d) along [001] directions 
are presented for W/N and W/S, respectively.\label{fig3}}
\end{figure}

\begin{figure*}[t!]
\epsfig{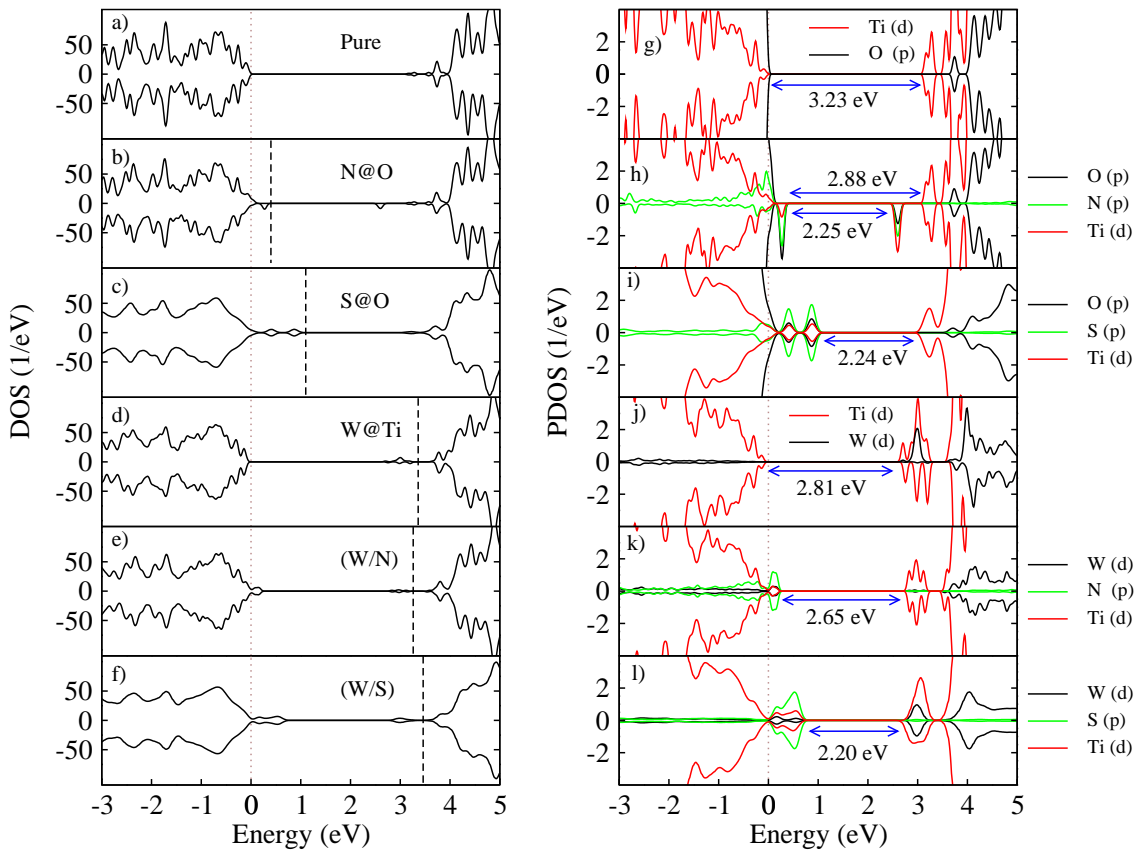}
\caption{Total (left pane) and projected (right pane) densities of states (DOS) of 
pure and, W and/or N(S) substituted anatase TiO$_2$, calculated with HSE06 
functional. Dashed lines indicate the Fermi energies. Dotted line denote the
VBM of pure TiO$_2$.
\label{fig4}}
\end{figure*}

\begin{figure*}[ht]
\epsfig{file=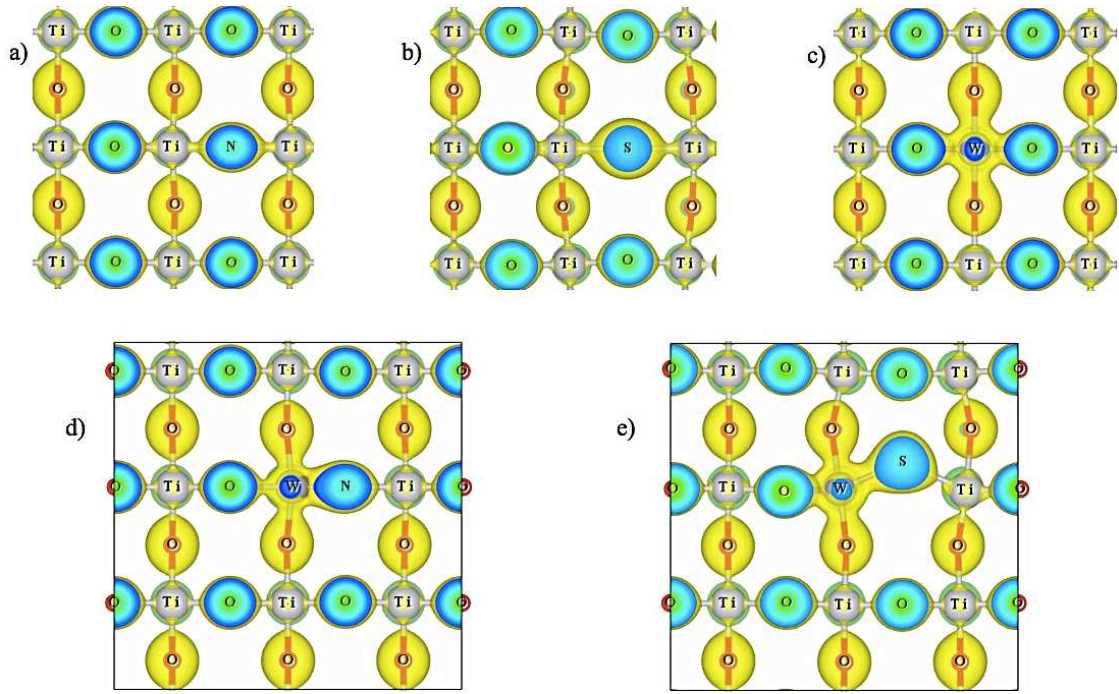,width=15cm}
\caption{Ground state total charge densities of (a) N-, (b) S-, (c) W-monodoped and
(d) W/N-, (e) W/S-codoped anatase TiO$_2$. These 3D plots have been cut through (001) 
plane slightly above the dopant species for visual convenience.\label{fig5}}
\end{figure*}

We considered W and N in the 108-atom cell (Fig.~\ref{fig1}) with different 
initial configurations, and found that the W-N pair shown in 
Fig.~\ref{fig3}a(c) is the lowest energy structure.  In agreement, 
Gao~\textit{et al.}\cite{Gao} reported the generation of N--W--O linkage to
be responsible for the increase in the visible light response. The interaction 
between W and N is much stronger with  HSE, giving a bond length of 1.78 
{\AA} which is significantly shorter than the GGA+U estimation of 1.85 
{\AA}.\cite{Long_WN2} W-N pair imposes small local distortions while 
maintaining the optimal anatase structural properties as pointed out by 
Kubacka~\textit{et al.}\cite{Kubacka}. For instance, T-N bond length 
being 2.00 {\AA} is slightly longer than 1.96 {\AA} of N@O case 
(in Fig~\ref{fig2}). Supportingly, Kubacka~\textit{et al.} emphasized 
that W presence in anatase lattice is likely to influence the Ti-N bond. 
Moreover, screened hybrid functional predicts lower formation energy 
for W/N codoping compared with previous theoretical studies. For example, 
we calculated E$_f$ as 5.87 eV under O-rich conditions. (Table~\ref{table2}) 
It  was estimated to be $\sim$8 eV by GGA+U method using a similar 
formulation.\cite{Long_WN2} 

Electronically, both W@Ti and N@O cells resulted in magnetic ground states with 
spin multiplicities of 0.8 and 1.0, respectively. The energy of the W 5$d$ 
donor level (in Fig~\ref{fig4}j) is compatible with that of the N 2$p$ acceptor 
level (in Fig~\ref{fig4}h). Hence, pairing of W with N in the lattice forms 
5$d$--2$p$ hybridization giving zero net magnetic moment as seen in 
Fig~\ref{fig4}k. This will, in turn, be effective in passivation of 
electron-hole recombination at dopant centers. 

Analysis of the Bader charges (in Table~\ref{table3}) reveals that N gets 
oxidized more by 0.7$e$ relative to N@O (see Fig.s~\ref{fig5}a,d). This charge 
is partially transferred from the nearest neighbor W (0.16$e$ by comparing with 
W@Ti) A large part is taken from the adjacent Ti atom such that its 
stoichiometric charge state is almost restored (by comparing Ti oxidation states 
in pure, N@O and W/N cases). In the presence of W-N pair, the oxidation state of 
the adjacent Ti is $+$2.82 (in Table~\ref{table3}). Briefly, the charge states 
of N and W get closer to formal oxidation values. This leads to a mutual 
passivation of N and W driven defect states. Indeed, the PDOS contribution of 
W 5$d$ in the isolated states below the CB of W@Ti disappears by mixing with 
N 2$p$ gap states.  Therefore, W/N codoping causes formation of isolated defect 
states 0.18 eV below the CB characterized by Ti 3$d$--N 2$p$ bonding. 
Consequently, HSE predicts a weak metallization for this nonstoichiometric 
codoping.

When W codopant pairs with N, the position of N driven state at the VBM of 
N@O essentially remains the same. The main difference, however,  is the 
disappearance of N 2$p$ acceptor level of  N@O. with the presence of W 5$d$ 
donor level. The CB edge shows similar characteristics with that of W@Ti. 
In Fig.~\ref{fig4}k, Ti 3$d$--N 2$p$ defect states lie 2.65 eV above the VB. 
This corresponds to an effective reduction of the band gap by 0.58 eV with 
respect to that of pure anatase. In previous experimental studies this narrowing 
was reported to be 0.55 eV.\cite{Kubacka,Kubacka1,Kubacka2} However, pure 
GGA and GGA+U studies found 0.2 and 1.11 eV gap narrowing values.
\cite{Long_WN1,Long_WN2} Partial compensation of deficient charge by 
W/N pairing was also confirmed to improve the photocatalytic efficiency 
supporting our results.\cite{Shen,Gao,Kubacka,Kubacka1,Kubacka2}   

\noindent\textbf{\emph{W/S codoping :}}
Monodoping of both S and W has been experimentally achieved to be effective in 
increasing the optical absorption efficiencies of TiO$_2$, suggesting the 
possibility of W/S codoping. This so that presence of S modifies the VB edge 
and brings defect states at the VBM by mixing with O $2p$ and Ti 3$d$ band 
states. In addition, shifting of the CBM and the 5$d$ donor levels are induced 
by W substitution. Therefore, as in the case of W/N, coexistence of S and W 
species is expected to have a cooperative effect on band gap narrowing.
Guided with this reasoning, we investigated the lowest energy configuration
of W and S substituted 108-atom anatase cell and found that W-S pairing 
as shown in Fig.~\ref{fig3}b,d is energetically more favorable than other
possibilities. For instance, relaxed geometry of non-pairing W-O-S 
configuration is 3.62 eV/cell higher in energy. W-S pair noticeably distorts 
the lattice relative to W/N case. However, the effect of this distortion 
remains local. The coordination of S with W and Ti increases W-S and Ti-S bond 
lengths. In fact, Ti-S bond being 2.19 {\AA} in S@O gets slightly larger by 
0.09 {\AA} by additional substitution of a W as shown in Fig.~\ref{fig3}d. Since 
relatively large W-S (2.25 {\AA}) and Ti-S (2.28 {\AA}) bonds do not fit 
in the anatase network, S atom relaxes out of the lattice site. This distortion 
results in increased formation energies with respect to those of W/N codoping 
as presented in Table~\ref{table2}, which suggests synthesis under O-rich 
conditions is more probable similar to the W/N.

Bader analysis reveals that S codopant gets more oxidized ($-$1.29$e$) with 
respect to S@O case ($-$1.17$e$). The charge states of the W ($+$4.20$e$) 
and of the adjacent Ti ($+$2.73$e$) indicate slightly less electron depletion 
from around them, particularly, relative to the W/N case. Passivative codoping 
effects of  the corresponding charge transfers can be seen in the DOS 
structures of the W/S system in Fig.s~\ref{fig4}f and \ref{fig4}l. For 
instance, as a result of W 5$d$--S 3$p$ hybridization, the PDOS peak of the W 
gets broadened inside the defect states relative to that of  the W@Ti case. In 
fact, these states form as a result of the mixing of W 5$d$--S 3$p$--Ti 3$d$ 
bonding states with largely dispersing Ti 3$d$ CB bands. As in the W-doped and 
the W/N cases, the Fermi energy is pinned above the W 5$d$ donor level similar 
to an oxygen vacancy situation.\cite{Janotti,Mete} Hybridization of orbitals is
not limited to these defect states at the CBM. S 3$p$ orbitals also mix with 
the VB continuum. Moreoever, a group of largely dispersing hybrid states 
(dominantly driven by S 3$p$ orbitals) form as a part of the VB edge. The shift 
of the VBM into higher energies is similar to the S monodoping case. This 
results in an effective band gap narrowing of 1.03 eV which is much larger 
than that of the W/N codoping. The apparent band gap value becomes 2.20 eV 
allowing optical absorption up to $\sim$564 nm. Partial passivation of 
W 5$d$ trap levels by S pairing might reduce electron-hole recombination 
rate. Also, W 5$d$ states are more delocalized relative to typical 3$d$ or 4$d$ 
orbitals and they strongly couple to the CB of TiO$_2$ which might help 
passivate the carrier recombination. Since impurity derived states at the
VBM and CBM are fully occupied, interband transitions are much more effective 
than localized state-to-band transitions, due to a much larger intensity.
Narrowing band gaps without creating isolated states are much more effective 
for photocatalytic activity.~\cite{Gai,HYu} In this sense, W/S codoping is 
predicted to drastically redshift the visible light absorption onset involving 
strong band-to-band and also band-to-state transitions, which, in turn, enhances 
the photocalaytic efficiencies.

\section{CONCLUSIONS}
We have systematically analyzed the atomic and electronic structures of N, S, 
W, W/N and W/S doping in anatase TiO$_2$ by the screened Coulomb potential 
approach. Our HSE results can be useful in explaining optical absorption 
spectra of these systems evidenced by many experiments. In this way, for the N 
doped-TiO$_2$, we identified an N 2$p$--O 2$p$ hiybrid state at the VBM and an 
N 2$p$ unoccupied trap level isolated 0.63 eV below the CB. This might serve
as an alternative explanation to solve some of the controversial experiment 
findings. Similarly, we are able to offer an understanding of the increased 
photocatalytic efficiencies observed for W/N codoping. The pair mixing of
N 2$p$ orbitals with the isolated W 5$d$ levels has been shown on the basis
of their DOS structures. W--N pairing results in a band gap narrowing of 0.58 
eV. As a trend, our HSE calculations might be useful in the interpretation of 
the passivative effect of such a pairing of a non-metal with a d-band metal. 
Then, we examined W/S codoping in anatase. By comparing modifications in the 
DOS structures we predict the origin of redshifts in the absorption light edge 
with W/N and W/S codoping systems. In fact, W/S codoping significantly reduces 
the optical absorption treshold which allows light harvesting in the large part 
of the visible range. Therefore, codoping of anatase TiO$_2$ with W and S is 
expected to achieve highly efficienct photocatalysis because the low lying 
photo-excitations involve transitions from S 3$p$ to Ti 3$d$ and W 5$d$ bands 
through an energy difference of 2.20 eV and the positons of these states are 
compatible with the redox potentials of water.

\begin{acknowledgments}
We acknowledge partial support from T\"{U}B\.{I}TAK, The Scientific and Technological Research 
Council of Turkey (Grant No. 110T394). Computational resources were provided by ULAKB\.{I}M,
Turkish Academic Network \& Information Center.
\end{acknowledgments}

\end{document}